\documentclass{aa501}
\usepackage{graphicx}
\def\arcsec{$^{\prime\prime}$}
\def\arcmin{$^{\prime}$}
\def\degrees{$^{\circ}$}

\begin{document}
\input psfig.sty

\date{Received ; accepted}

\title{Radio galaxies and magnetic fields in A514}
\author{F. Govoni\inst{1,2} \and G.\ B. Taylor\inst{3}
\and D. Dallacasa\inst{1,2} \and L. Feretti\inst{2} \and
G. Giovannini\inst{2,4}}

\offprints{F. Govoni (fgovoni@ira.bo.cnr.it)}
\institute{
Dipartimento di Astronomia, Univ. Bologna, Via Ranzani 1, I--40127 Bologna, Italy
\and Istituto di Radioastronomia -- CNR, via Gobetti 101, I--40129
Bologna, Italy
\and 
National Radio Astronomy Observatory, Socorro, NM 87801, USA
\and Dipartimento di Fisica, Univ. Bologna,  
Via B. Pichat 6/2, I--40127 Bologna, Italy
}

\abstract{
A514 contains six extended and polarized radio sources located at
various projected distances from the cluster center.
Here we present
a detailed study of these six radio sources in total intensity and
polarization using the Very Large Array at 3.6 and 6 cm.  Since
the radio sources sample different lines of sight across the cluster,
an analysis of the Faraday Rotation Measures (RMs) provides
information on the strength and the structure of the cluster magnetic
field.  These sources show a decreasing Faraday Rotation Measure with
increasing distance from the cluster center.  We estimate the strength
of the magnetic field to be $\sim$3-7 $\mu$G in the cluster center. 
From the RM structure across the
stronger and more extended sources we estimate the coherence length of
the magnetic field to be about 9 kpc at the cluster center.
}

\maketitle

\section{Introduction}

The study of the magnetic field in clusters of galaxies 
is relevant to understand the physical conditions and energetics 
of the intra-cluster medium (ICM),
the cluster formation and evolution,
the origin of the intergalactic 
magnetic field itself (primordial, Olinto 1997; injected 
into the ICM from galactic winds or AGNs, Kronberg et al. 1999; 
V\"olk \& Atoyan 1999), and
its consequences on primordial star formation 
(Pudritz \& Silk 1989).

The presence of a synchrotron radio halo in a cluster of galaxies is
a direct evidence for the presence of substantial magnetic fields
in the cluster.
Radio halos are produced by diffuse, non-thermal emission
associated with the ICM rather than with a particular galaxy of 
the cluster, and have sizes of the order of 1 Mpc (e.g. Coma C,
Giovannini et al. 1993).

So far, the studies of the magnetic fields in clusters
of galaxies, obtained from different methods,
give somewhat discrepant measurements for the field strengths.

Assuming minimum energy content, ({\it i.e.}, 
nearly energy equipartition between the magnetic field and the
relativistic particles), measurements of the strength of the
magnetic fields in some radio halos have been made, and reveal
magnetic field strengths of about 0.1-1 $\mu G$ 
(Feretti \& Giovannini 1996).  
In the Coma cluster, the equipartition magnetic field is
$\simeq 0.4h_{50}^{2/7}\mu G$ (Giovannini et al. 1993).  

Estimates by Fusco-Femiano et al. (1999), obtained by comparing 
the diffuse radio emission
with the hard X-ray emission in the Coma cluster 
(assumed to arise from Inverse Compton scattering of the 
cosmic microwave background of the relativistic electron 
responsible of the radio emission)
give a field of 0.15 $\mu G$.

Measurements of the magnetic field strength can also be
determined in conjunction
with X-ray observations of the hot gas, through the study of the
Faraday Rotation Measure (RM) of radio sources located 
inside or behind the cluster.

Strong magnetic fields, from $\sim$5 up to the
value of 30 $\mu$G have been found in ``cooling flow'' 
clusters ({\it e.g.} 3C~295, Perley \& Taylor 1991 and 
Allen et al. 2001; Hydra A, Taylor \& Perley 1993) 
where extremely large Faraday rotations have been revealed, suggesting that the
generation of very strong ICM magnetic fields may be connected with
the cooling flow process (Soker \& Sarazin 1990; Godon et al. 1998).

On the other hand, significant magnetic fields have also been 
detected in clusters without a cooling flow: the measurements of 
Faraday rotation of polarized radiation through the hot ICM 
leads to a magnetic field of 2-6 $\mu G$ 
in the Coma cluster (Kim et al. 1990, Feretti et al. 1995). 
Similarly, the presence
of a magnetic field was detected in the cluster A119 (Feretti et al. 1999)
through the analysis of
the Faraday Rotation in three extended radio galaxies located at different
distances from the cluster center.
We found that the RM decreases with the distance from the cluster
center, and it is consistent with a field strength of
about 5-10 $\mu G$ at the cluster center, depending on the tangling scale.
These values are quite similar to the field strength of about 
5 $\mu$G found in the cluster containing the radio source 
3C129 (Taylor et al. 2001) by similar analysis.  
From a statistical study of several clusters, Clarke et al. (2001)
obtained magnetic fields of 4-8 $\mu G$, confirming the previous
findings.

By studying the spatial 
distribution of the RMs we can also estimate
the coherence length of the magnetic field.
The ICM magnetic field can be tangled on scales much smaller
than the typical galaxy size.
Crusius-W\"atzel et al. (1990), studying the depolarization in 5 strong double
sources, found tangling on scales of 1-4 kpc. This is confirmed
also by the results on the Coma, A119 and 3C129 clusters 
(Feretti et al. 1995, Feretti et al. 1999, Taylor et al. 2001).

We note that the field strength measured from the RM 
may be influenced 
by the presence of filamentary structures (Eilek 1999) and by local
turbulences in the ICM induced by the host galaxy motion
within the cluster. 
Therefore the value estimated by RM arguments
is generally higher than the
 equipartition magnetic field, 
which refers to the ``average'', uniformly distributed
field permeating the cluster (see also Goldsmith \& Rephaeli 1993).

Here we present the analysis of the magnetic field in the cluster
Abell 514, which is characterized by a clumpy 
and elongated X-ray surface brightness typically found in clusters
associated with a merging process.

No radio halo is reported in the literature for this cluster.
We analyzed the image taken from the 
NRAO VLA Sky Survey (NVSS, Condon et al., 1998),
with a resolution of 45$''$ and 
a sensitivity of 0.45 mJy/beam (see contours in Fig. \ref{fig_RadX}),
in a region of about 5$'$ around the cluster center.
No evidence for extended-diffuse low surface brightness
emission at the noise level was found. 
Furthermore, we measured the total flux
density in the aforementioned region and then we subtracted
the contribution of each individual source. The resulting
difference is consistent with the absence of diffuse emission.

However A514 contains six moderately strong, extended, and
polarized sources located inside or behind the cluster
and at different projected distances from the cluster center.
Therefore this cluster is well suited for the analysis 
of the magnetic field through the study of the Faraday Rotation 
Measure.

We give a brief description of the cluster in \S2. 
In \S3 we describe the radio observations.  In \S4 we present the
total intensity and polarization images for all the sources, and in
\S5 the X-ray image and the important X-ray parameters are analyzed.
Finally in \S6 and \S7 we discuss the Rotation Measure results and
the presence of magnetic fields.  We assume H$_0$=50 km s$^{-1}$
Mpc$^{-1}$ and q$_0$=0.5 throughout the paper.  At the distance of
A514 (luminosity distance $D_L\simeq435.5$~Mpc) 1$''$ corresponds to
1.84 kpc.

\section{Abell 514}

A514 (z=0.0714 Fadda et al. 1996)
is of richness class 1 (Abell et al. 1989).
It is classified as BM type II-III (Bautz \& Morgan 1970)
and as Rood-Sastry type F (Rood \& Sastry 1971).

\begin{figure*}
\begin{center}
\includegraphics[width=13cm, angle=0]{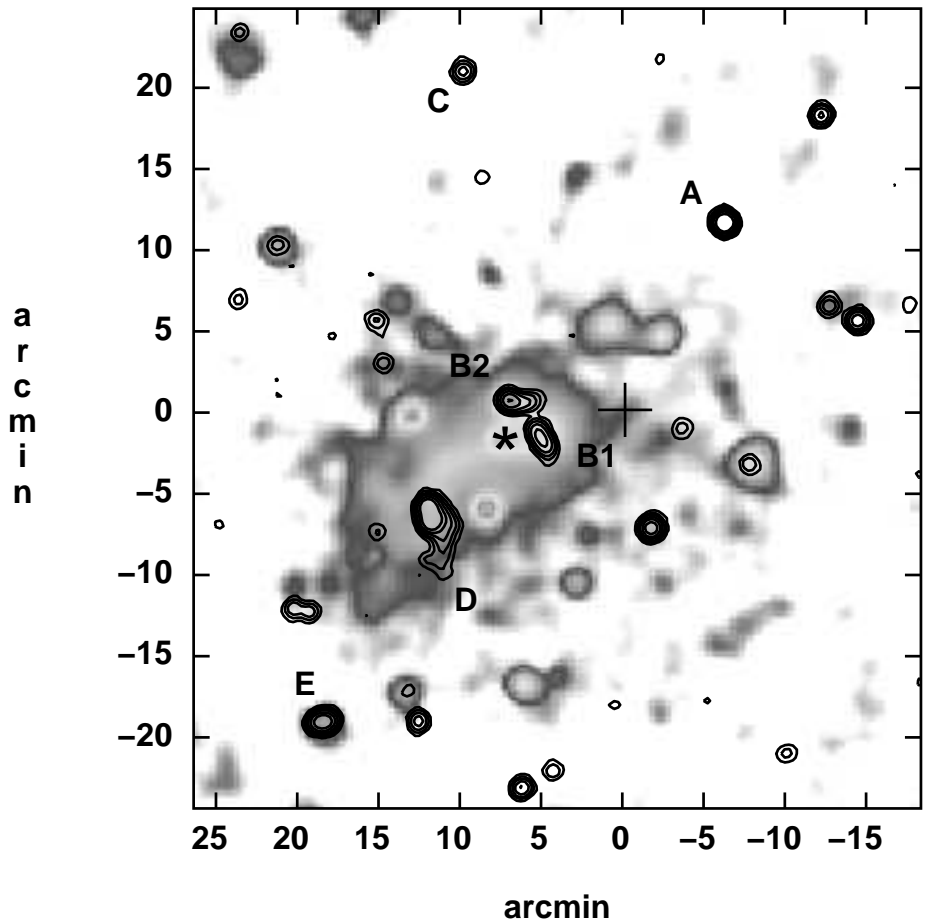}
\end{center}
\hfill
\caption[]{The 1.4 GHz radio image obtained from the NVSS (contour)
superimposed to the X-ray image of Abell 514 taken from ROSAT 
PSPC archive (grey scale). 
Radio contour levels are 2, 5, 10, 20, 40, and 80 mJy/beam.
The radio image has an angular resolution of 45$''$.
The ROSAT image has an angular resolution of 25$''$ (FWHM) and 
has been smoothed with a Gaussian of $\sigma=30''$.
The reference point of the image (indicated with a cross in the figure) 
was chosen to be the optical center of the cluster 
indicated by Abell et al. (1989).
The X-ray center of the cluster is indicated with a star.}
\label{fig_RadX}
\end{figure*}

Fig. \ref{fig_RadX} shows the radio image at 1.4 GHz obtained from the
NVSS, overlaid on the X-ray image
obtained by us from the ROSAT PSPC archive. The reference point of the image
(indicated with a cross in the figure) was chosen to be the optical
center of the cluster indicated by Abell et al. (1989) at
RA(J2000)=04$^h$47$^m$40$^s$, DEC(J2000)=$-$20\degrees 25.7\arcmin.
The X-ray surface brightness is rather smooth, without any 
outstanding peak. We adopted as  X-ray center of the cluster
the highest X-ray brightness peak not related to any point-like emission,
at position RA(J2000)=04$^h$48$^m$13$^s$, 
DEC(J2000)=$-$20\degrees 27\arcmin 18\arcsec.
The X-ray center of the cluster appears shifted 
to the East with respect the optical center of about 8\arcmin.
In Fig. \ref{fig_RadX} the X-ray center of the cluster 
is indicated with a star. 
In \S5 we will discuss in greater detail the X-ray emission
of A514 and its implications on the parameter estimates.

Here we examine multi-wavelength Very Large Array 
(VLA\footnote{The National Radio Astronomy Observatory is operated by
Associated Universities, Inc., under cooperative agreement with the National
Science Foundation.}) observations
of six radio sources within or behind cluster Abell 514
indicated by labels in Fig. \ref{fig_RadX}.
The two radio galaxies which lie close to 
the cluster center (B1, B2) show a
narrow-angle-tailed structure, the radio galaxy D is a tailed, while the
other two radio galaxies (A, C) are FRII type radio galaxies.  The 
source E located in the cluster periphery is classified as a quasar.

We derive the Rotation Measures (RM) of the radio sources, which,
together with the X-ray data, provide information about the magnetic 
field in the cluster.
 
\section{Radio Observations}

\begin{table*}
\caption{Pointing and VLA observations of A514.}
\begin{flushleft}
\begin{tabular}{ccccccccc}
\hline 
\noalign{\smallskip}
Name & Other Name       & Label      &RA      & DEC   &  Frequency  & Config. & Date & Duration     \\
     &                  &            &(J2000) &(J2000)& (MHz)       &         &      & (Hours)        \\
\noalign{\smallskip}
\hline
\noalign{\smallskip}
J0447$-$2014   & PKS $0445-203$ &   A    & 04:47:12.8 & -20:13:58.0  & 4535/4885   & BnA,CnB & Oct.99,Mar.00    & 1.2  \\
               &                &            &            &          & 8085/8465   & BnA,CnB & Oct.99,Feb.00    & 1.2  \\
J0448$-$2026   &                &   B1   & 04:48:03.0 & -20:26:31.0  & 4535/4885   & BnA,CnB & Oct.99,Mar.00    & 3.0  \\
               &                &            &            &          & 8085/8465   & BnA,CnB & Oct.99,Feb.00    & 2.6  \\
J0448$-$2025   &                &   B2   & 04:48:10.5 & -20:24:56.0  & 4535/4885   & BnA,CnB & Oct.99,Mar.00    & 3.0  \\
               &                &            &            &          & 8085/8465   & BnA,CnB & Oct.99,Feb.00    & 2.6  \\ 
 J0448$-$2005   &                & C    & 04:48:21.7 & -20:04:48.0   & 4535/4885   & BnA,CnB & Oct.99,Mar.00 & 1.2  \\
       &                &            &            &                  & 8085/8465   & BnA,CnB & Oct.99,Feb.00 & 1.2  \\
 J0448$-$2032    & PKS $0446-206$ & D     & 04:48:30.4 & -20:31:49.0 & 4535/4885   & BnA,CnB & Oct.99,Mar.00 & 3.0  \\
       &                &            &            &                  & 8085/8465   & BnA,CnB & Oct.99,Feb.00 & 2.6  \\
 J0448$-$2044    & PKS $0446-208$ & E     & 04:48:59.0 & -20:44:51.0 & 4535/4885   & BnA,CnB & Oct.99,Mar.00 & 1.2  \\
       &                &            &            &                  & 8085/8465   & BnA,CnB & Oct.99,Feb.00 & 1.2  \\
\noalign{\smallskip}
\hline
\multicolumn{9}{l}{\scriptsize Col. 1: Source name; Col. 2: Other name; 
Col. 3: Label name; Col. 4, Col. 5: Pointing position (RA, DEC); 
Col. 6: Observation frequencies;}\\
\multicolumn{9}{l}{\scriptsize Col. 7: VLA configuration; 
Col. 8: Dates of observation; Col. 9: Duration of observation.}\\ 
\label{observ}
\end{tabular}
\end{flushleft}
\end{table*}

The radio sources were observed with the VLA at two frequencies
within the 6 cm band and at two frequencies within the 3.6 cm band
with a bandwidth of 50 MHz.
In Table \ref{observ} we provide details of the observations.
We note that given the southern declination of A514, the VLA
was used in hybrid configurations (BnA,CnB) 
in order to have roughly circular restoring beams.

The phase calibration was performed using the secondary calibrator
{\tt 0457-234}. The flux-density scale and the absolute 
polarization position angle were calibrated by
observing 3C48.  The instrumental polarization of the
antennas was corrected using {\tt 0457-234} and also {\tt 0530+135} 
in CnB configuration observed over a wide range of parallactic angles.

Images in all Stokes parameters were produced with the NRAO AIPS
package following the standard procedures.  Self calibration was applied
to minimize the effect of amplitude and phase uncertainties of
atmospheric and instrumental origin. The (u,v) data at the same frequencies
but from different configurations were first handled separately and then
added together.

The images of the polarized intensity $P=(Q^2+U^2)^{1/2}$, the
degree of polarization $m=P/I$ and the position angle of polarization
$\Psi=0.5\tan^{-1}(U/Q)$ were derived from the I, Q and U images.
Total intensity images have been produced by averaging the two
frequencies in the same band (4535/4885 MHz; 8085/8465 MHz)
while U and Q images have been obtained for each 
frequency separately. 

\section{Total intensity and polarization images}

In Table \ref{mappeI} we give the parameters
for the total intensity images at 1.6$''$ ($\simeq$ 3 kpc) resolution. 
The total flux density was estimated after the primary beam correction.
In all the radio images presented in this work
contours are total intensity while vectors represent 
the orientation of the projected E-field and their 
length is proportional to the fractional polarization.
In the polarization images,
points with error in fractional polarization greater than $10\%$ 
were clipped.
In the following we present the individual sources.

\subsection{J0447-2014 (A514A)}

\begin{figure*}
\resizebox{18 cm}{!}{\includegraphics {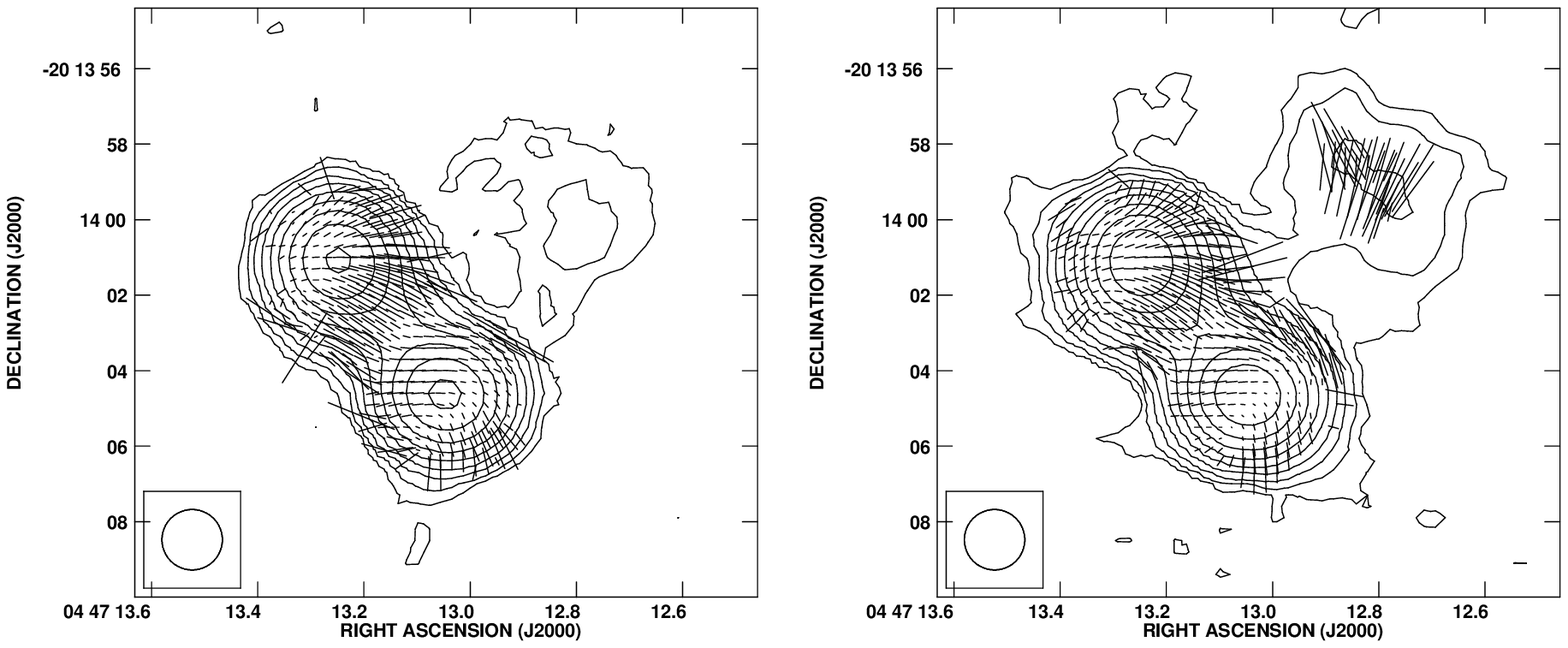}}
\hfill
\caption[]{Left: Image of {\tt J0447-2014} (A514A) 
at 3.6 cm with an angular resolution of $1.6''\times1.6''$. 
The rms noise level is 0.027 mJy/beam. The peak is 29.6 mJy/beam.
Right: Image of {\tt J0447-2014} (A514A) 
at 6 cm with an angular resolution of $1.6''\times1.6''$.
The rms noise level is 0.028 mJy/beam.
The peak is 47.2 mJy/beam.
In both images contour levels are:
$-$0.1, 0.1, 0.2, 0.4, 0.8, 1.6, 3.2, 6.4, 12.8, and 25.6 mJy/beam.
The lines represent the orientation of the
electric vector (E-field) and are proportional in length to the 
fractional polarization ($1''\simeq11\%$).
}
\label{fig_A514}
\end{figure*}

\begin{figure}
\resizebox{6.5 cm}{!}{\includegraphics {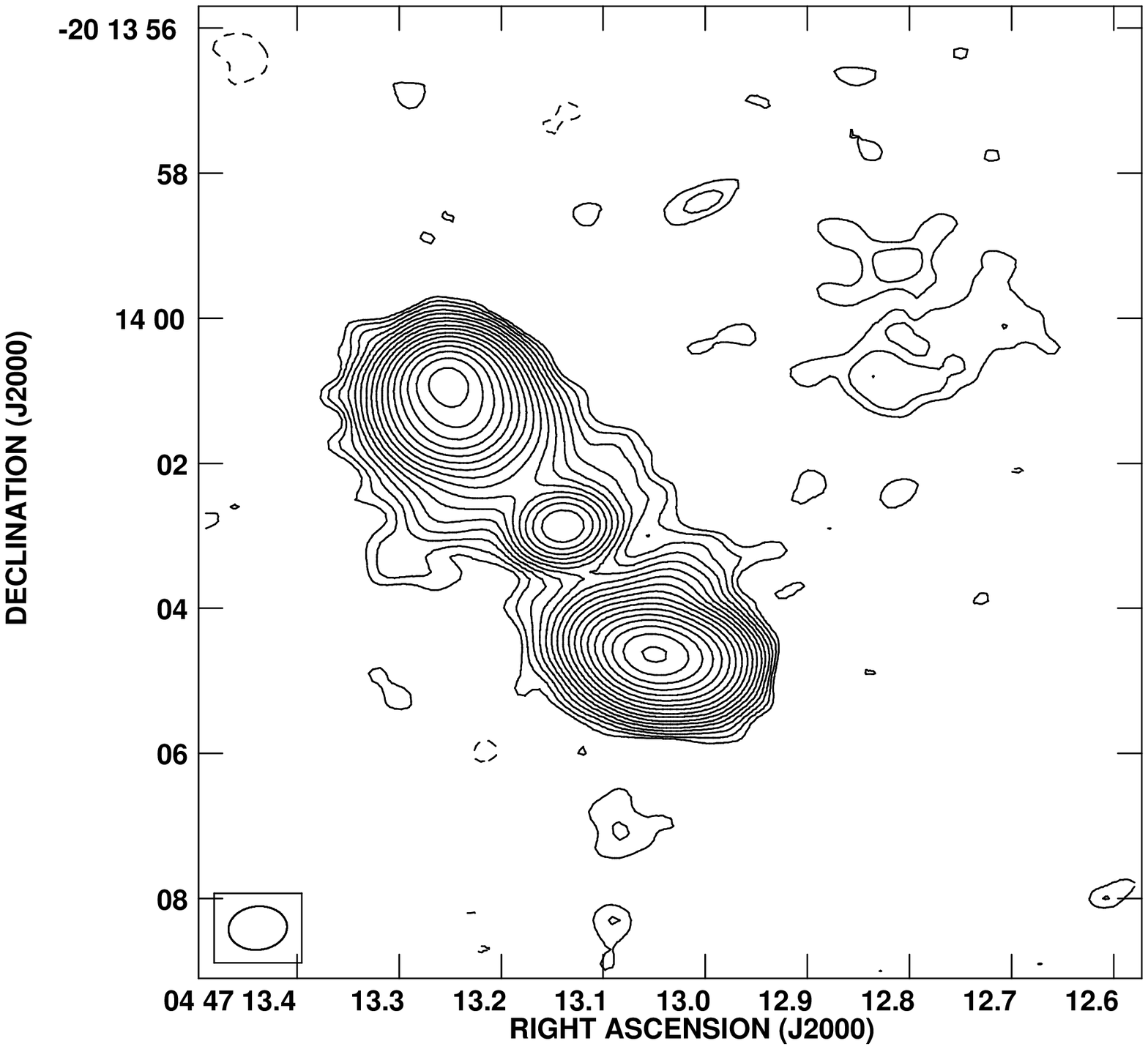}}
\hfill
\caption[]{Image of {\tt J0447-2014} (A514A) at 3.6 cm with an angular resolution 
of $0.8''\times0.6''$.
Contour levels are: $-$0.07, 0.07, 0.1, 0.14, 0.20, 0.28, 0.40, 0.56, 0.79,
1.12, 1.58, 2.24, 3.17, 4.48, 6.34 mJy/beam.
The rms noise level is 0.023 mJy/beam. 
The peak is 19.1 mJy/beam.}
\label{fig_A514bis}
\end{figure}

In the Digitized  Sky Survey (DSS) image from the UK Schmidt Telescope
there is a very faint 
optical object coincident with the radio source.
No redshift is available, 
but the object appears fainter by about 2 magnitudes
than the other cluster members.
Its peripheral position and optical properties suggest
that it is a background source.

This radio source was observed by Owen et al. (1996) 
in a VLA Survey of radio source in Abell clusters.
They excluded this object from their complete sample of radio galaxies
in Abell Cluster because the source is at a projected distance 
larger than 0.3 Abell radii.
Their radio observations were made with the VLA at 1.4 GHz 
in B and C Array. They measured a total flux density of 270 mJy
obtained by fitting the unresolved source with a Gaussian model.
 
Our observations resolve the radio source.
In Fig. \ref{fig_A514} we show the radio images at 3.6 cm and 
6 cm with an angular resolution of 1.6$''$.
The maximum projected angular size of the source is about $12''$.
At both frequencies, the radio source shows a double 
structure. An extension in the North-West
side is visible in the images.

The source is polarized at 3.6 cm and 6 cm, with similar
values of the polarization percentage.
The fractional polarization is $\simeq5\%$ in the northern 
lobe and $\simeq 4\%$
in the southern lobe. 
The extended emission in the North-West is  polarized at 6 cm
(about 10$\%$) while at 3.6 cm the sensitivity is not 
enough to reveal any significant polarized emission and we only
set an upper limit of about 13$\%$ to the fractional polarization.

\begin{figure*}
\resizebox{18 cm}{!}{\includegraphics {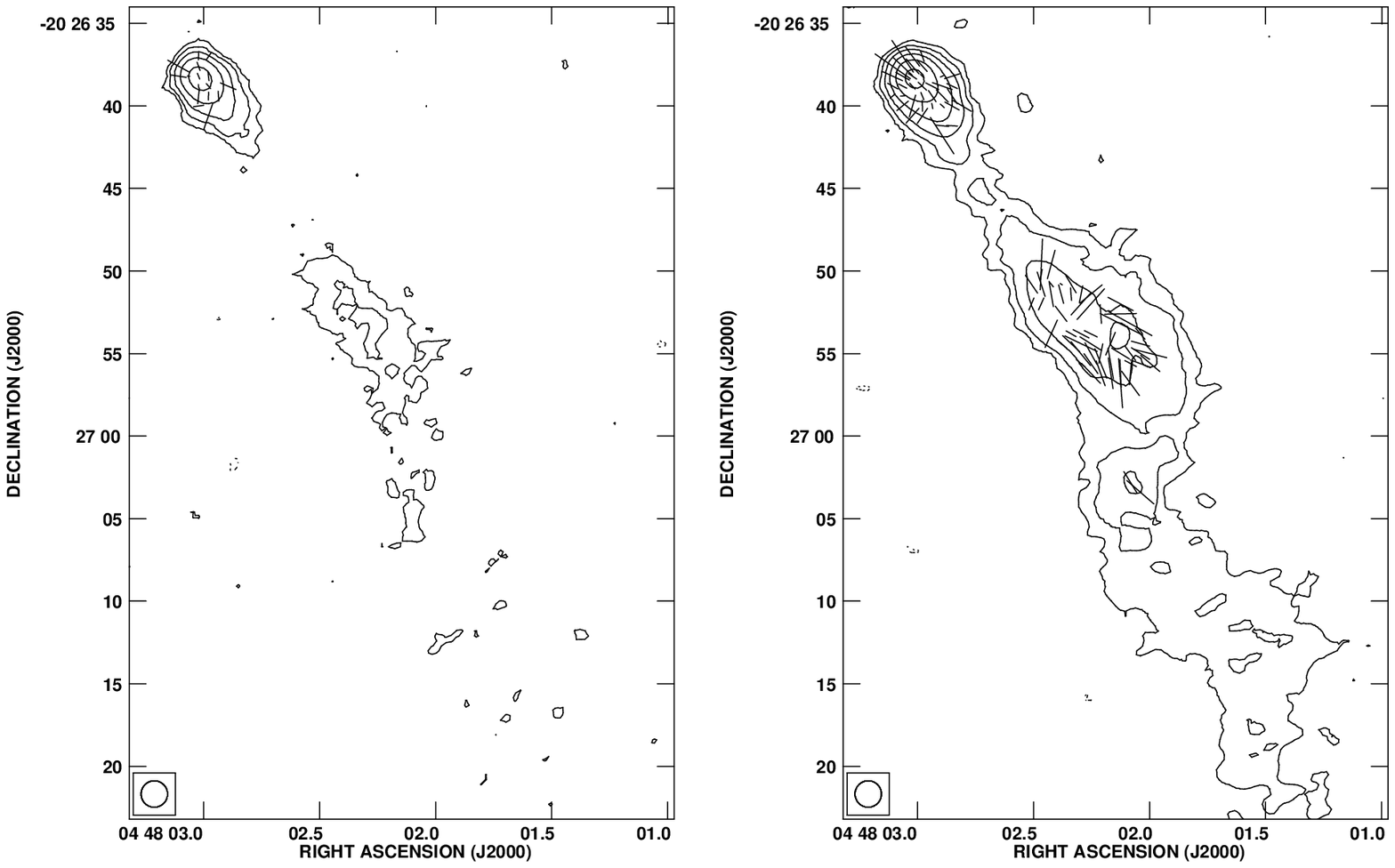}}
\hfill
\caption[]{Left: Image of {\tt J0448-2026} (A514B1) 
at 3.6 cm with an angular resolution of $1.6''\times1.6''$.
The rms noise level is 0.019 mJy/beam.
The peak is 4.1 mJy/beam.
Right: Image of {\tt J0448-2026} (A514B1) at 6 cm 
with an angular resolution 
of $1.6''\times1.6''$.
The rms noise level is 0.018 mJy/beam.
The peak is 8.2 mJy/beam.
In both images contour levels are:
$-$0.06, 0.06, 0.12, 0.24, 0.48, 0.96, 1.92, 3.84, and 7.68 mJy/beam.
Superimposed lines represent the orientation of the
electric vector (E-field) and are proportional in 
length to the fractional polarization ($1''\simeq9\%$).}
\label{fig_A514B1}
\end{figure*}

In  Fig. \ref{fig_A514bis}
we present the same source at 3.6 cm with the highest 
angular resolution. 
The image shows the unresolved nucleus at position: 
RA(2000)=04$^h$47$^m$13.15$^s$, DEC(2000)=$-$20\degrees 14\arcmin 03\arcsec.
For this component we estimate a spectral 
index $\alpha \simeq 0.7$ (where $\alpha$ is defined as 
$S(\nu)\propto \nu^{-\alpha}$); at this resolution,
this quite steep spectrum
in the compact component could be due to the contamination of 
other structures in the source. 
We found a spectral index $\alpha \simeq$ 0.9
and $\alpha \simeq$ 0.8 respectively in the northern and in the southern
lobe.

\begin{figure*}
\resizebox{12 cm}{!}{\includegraphics {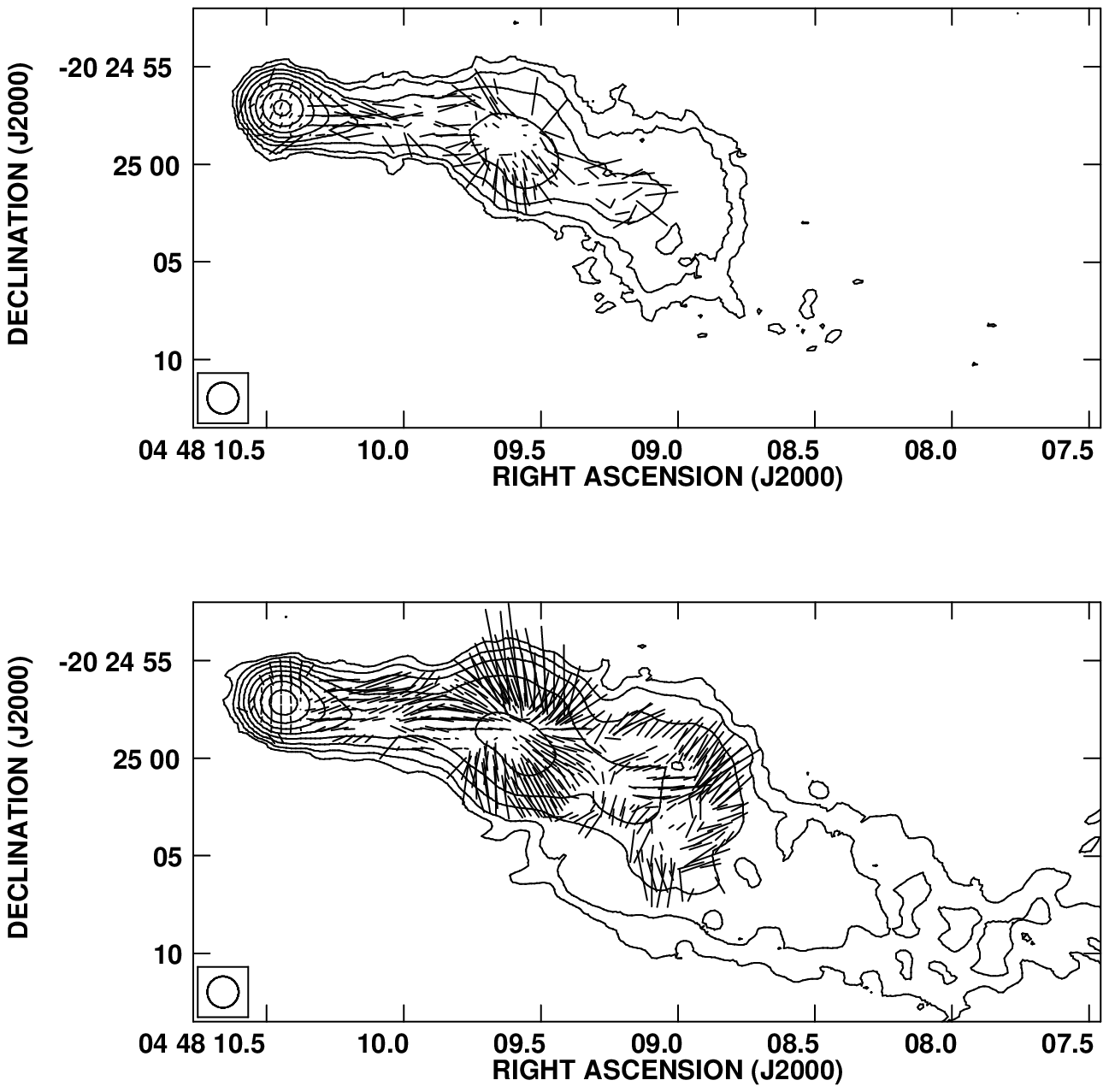}}
\hfill
\caption[]{Top: Image of {\tt J0448-2025} (A514B2) 
at 3.6 cm with an angular resolution of $1.6''\times1.6''$.
The peak is 9.1 mJy/beam.
Bottom: Image of {\tt J0448-2025} (A514B2) at 6 cm 
with an angular resolution of $1.6''\times1.6''$.
The peak is 11.4 mJy/beam.
In both images contour levels are:
$-$0.06, 0.06, 0.12, 0.24, 0.48, 0.96, 1.92, 3.84, and 7.68 mJy/beam
and the rms noise level is 0.017 mJy/beam.
Superimposed lines represent the orientation of the
electric vector (E-field) and are proportional in length to the 
fractional polarization ($1''\simeq10\%$).}
\label{fig_A514B2}
\end{figure*}

\begin{table*}
\caption{Parameters of total intensity radio images, with 1.6$''$ resolution.}
\begin{tabular} {ccccc} \hline
Source    & $\lambda$      & $\sigma$(I)     &Peak brightness  &Flux density    \\
          & (cm)           & (mJy/beam)      &(mJy/beam)&(mJy)                   \\ 
          &                &                 &          &                     \\\hline
$J0447-2014$  &  3.6       & 2.7$\times10^{-2}$  & 29.6 &     79.2              \\
 $''$         &  6         & 2.8$\times10^{-2}$  & 47.2 &    131.0              \\

$J0448-2026$  &  3.6       & 1.9$\times10^{-2}$  &  4.1 &     6.1       \\ 
 $''$         &  6         & 1.8$\times10^{-2}$  &  8.2 &    23.5       \\

$J0448-2025$  &  3.6       & 1.7$\times10^{-2}$  &  9.1 &     25.3       \\ 
 $''$         &  6         & 1.7$\times10^{-2}$  & 11.4 &     51.4       \\ 

$J0448-2005$  &  3.6       & 2.4$\times10^{-2}$  &  0.9 &      7.6       \\ 
 $''$         &  6         & 2.4$\times10^{-2}$  &  1.6 &     14.5       \\

$J0448-2032$  &  3.6       & 1.8$\times10^{-2}$  &  7.0 &     80.7       \\ 
 $''$         &  6         & 2.0$\times10^{-2}$  &  6.4 &    190.5       \\

$J0448-2044$  &  3.6       & 3.0$\times10^{-2}$  & 48.9 &     70.6       \\
 $''$         &  6         & 2.6$\times10^{-2}$  & 65.3 &    105.9       \\  \hline 
\noalign{\smallskip}
\multicolumn{5}{l}{\scriptsize Col. 1: Source name; Col. 2: Observation wavelength;}\\
\multicolumn{5}{l}{\scriptsize Col. 3: RMS noise; Col. 4: Peak brightness; 
Col. 5: Flux density.}\\ 
\label{mappeI}
\end{tabular}
\end{table*}

\subsection{J0448-2026 (A514B1)}

A 1.4 GHz image is presented in Owen et al. (1993) who measured a
total flux density of 95 mJy.
They considered this galaxy to be a member of the cluster.

In Fig. \ref{fig_A514B1} we show the total intensity and polarization 
images at 3.6 cm and at 6 cm
with an angular resolution of 1.6$''$.
The source shows a Narrow-Angle Tail (NAT) structure
and in projection it is located near the cluster center.
In our 6 cm image the source has a size of 66$''$ ($\simeq$ 120 kpc).
At 3.6 cm the head is readily visible, while most of the low brightness
emission in the tail is below the sensitivity of the present observations.
The mean spectral index is $\alpha \simeq 0.9$ in the
head while it is steeper ($\alpha \simeq 1.8$) in the tail.
At both wavelengths the head is $\sim4\%$ polarized, while the tail is polarized ($\simeq 5\%$)
at 6 cm but not detected in polarized flux at 3.6 cm.
In the tail, at 3.6 cm, we quote an upper limit of 26$\%$ to the fractional
polarization. Due to its low polarization this source was not used
to study the rotation measure.

\subsection{J0448-2025 (A514B2)}

Slee et al. (1994) identified the host galaxy as a D galaxy.
The radio source morphology is that of a head tail source.
The source was also studied by Owen et al. (1993, 1997) 
to be a member of the cluster.
A 1.46 GHz image is shown in Owen et al. (1993) who calculated
a total flux density of 119 mJy at this frequency.

In Fig. \ref{fig_A514B2} we show the total intensity and
polarization images obtained at 3.6 cm and at 6 cm. 
In both images the morphology of
the radio galaxy is well defined as tailed elongated to the West.
At 6 cm the total intensity radio emission reaches an extension of 50$''$,
while at the same wavelength the polarized emission 
is detected only out to 25$''$ from the core.
At both wavelengths the percentage of polarization is $\sim$4\%
in the head.
The polarization along the tail is $\sim$8\% at 6 cm.
The mean spectral index is $\alpha \simeq$ 0.5 in the head
while it becomes steeper in the brightest part of
the tail, $\alpha \simeq$ 1.2.

\subsection{J0448-2005 (A514C)}

\begin{figure*}
\resizebox{18 cm}{!}{\includegraphics {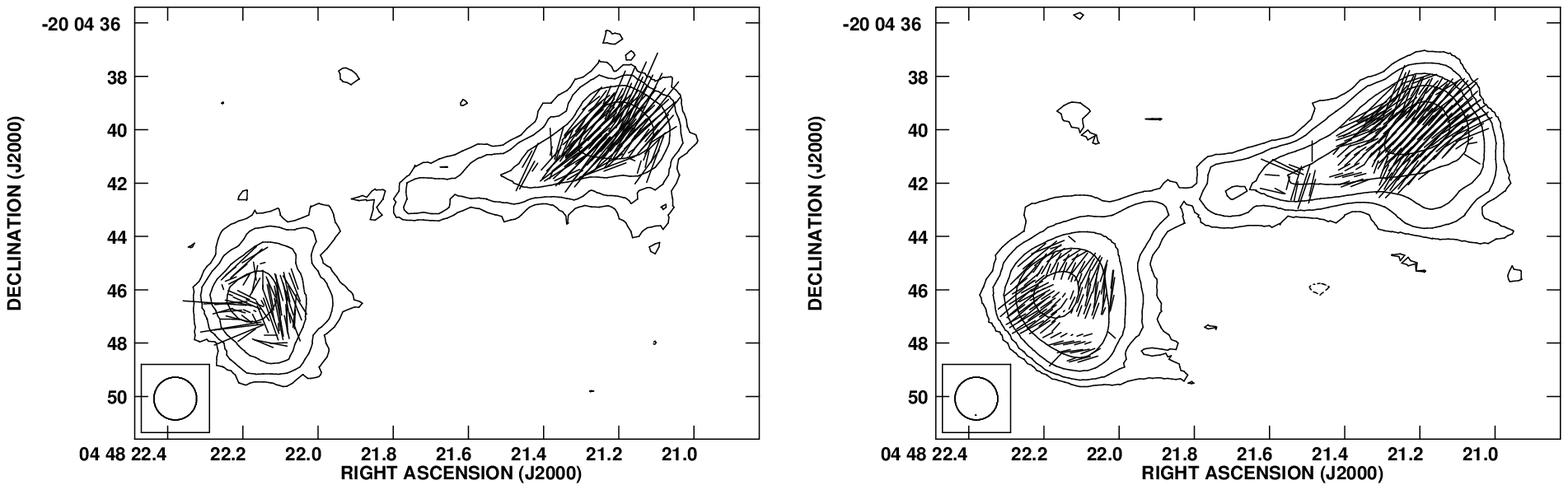}}
\hfill
\caption[]{Left: Image of {\tt J0448-2005} (A514C) at 
3.6 cm with an angular resolution of $1.6''\times1.6''$.
The peak is 0.9 mJy/beam.
Right: Image of {\tt J0448-2005} (A514C) at 6 cm with an angular resolution 
of $1.6''\times1.6''$.
The peak is 1.6 mJy/beam.
In both images contour levels are:
$-$0.07, 0.07, 0.14, 0.28, 0.56, 1.12, 2.24, 4.48, and 8.96 mJy/beam
and the rms noise level is 0.024 mJy/beam.
Superimposed lines represent the orientation of the
electric vector (E-field) and are proportional in length to the 
fractional polarization ($1''\simeq11\%$).}
\label{fig_A514C}
\end{figure*}

No reference for this radio source has been found in the literature
and no optical identification is seen on the DSS. Since the
host galaxies of radio sources are typically bright ellipticals, 
we expect for the parent galaxy an absolute R
magnitude at least of $M_R=-20.0$  (Auriemma et al. 1977) 
which corresponds to an apparent magnitude
$m_R=18.2$ at the cluster redshift.  A galaxy with this apparent
magnitude would be readily seen in the DSS if the galaxy belongs to
the cluster.  Due to the absence of the optical identification and to
the peripheral location of the source with respect to the cluster
center, we suggest that this object is a background source.

\begin{figure}
\resizebox{9 cm}{!}{\includegraphics {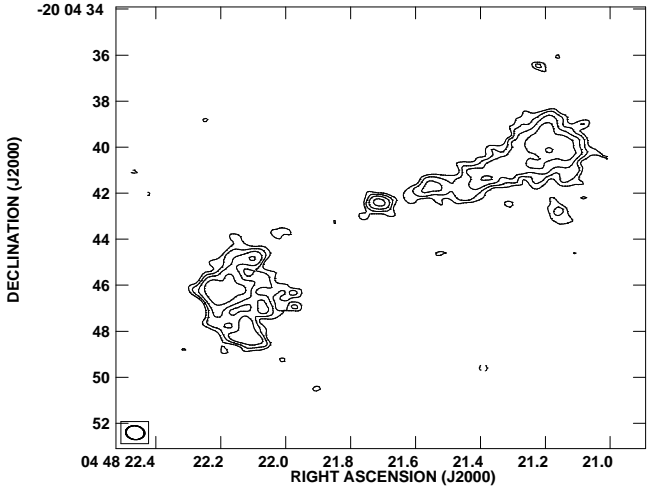}}
\hfill
\caption[]{Image of {\tt J0448-2005} (A514C) at 3.6 cm with an angular resolution 
of $0.8''\times0.6''$.
Contour levels are: $-$0.07, 0.07, 0.10, 0.14, 0.20, 0.28 mJy/beam.
The rms noise level is 0.02 mJy/beam.
The peak is 0.29 mJy/beam.}
\label{fig_A514CC}
\end{figure}

In Fig. \ref{fig_A514C} 
we show the images of {\tt J0448-2005} at 3.6 cm and at 6 cm 
with an angular resolution of 1.6$''$.
The maximum projected angular size of the source is about 22$''$.
The radio source morphology is similar to FR-II type radio galaxies
with the brightest regions at the outer edges of the images. 
The full resolution 3.6 cm image (Fig. \ref{fig_A514CC}) reveals also a 
compact component near the center of the radio source.
The spectral index in this region is $\simeq 0.4$ suggesting that it
could harbor the source core.
In the eastern lobe the mean spectral index is $\simeq 1.0$
while  in the western lobe it is $\simeq 1.2$.
No polarized emission is detected from the core while the two lobes 
are strongly polarized at both wavelengths -- the western lobe is 14$\%$ polarized, and the
eastern lobe is 7$\%$ polarized.
  
\subsection{J0448-2032 (A514D)}

\begin{figure*}
\resizebox{18 cm}{!}{\includegraphics {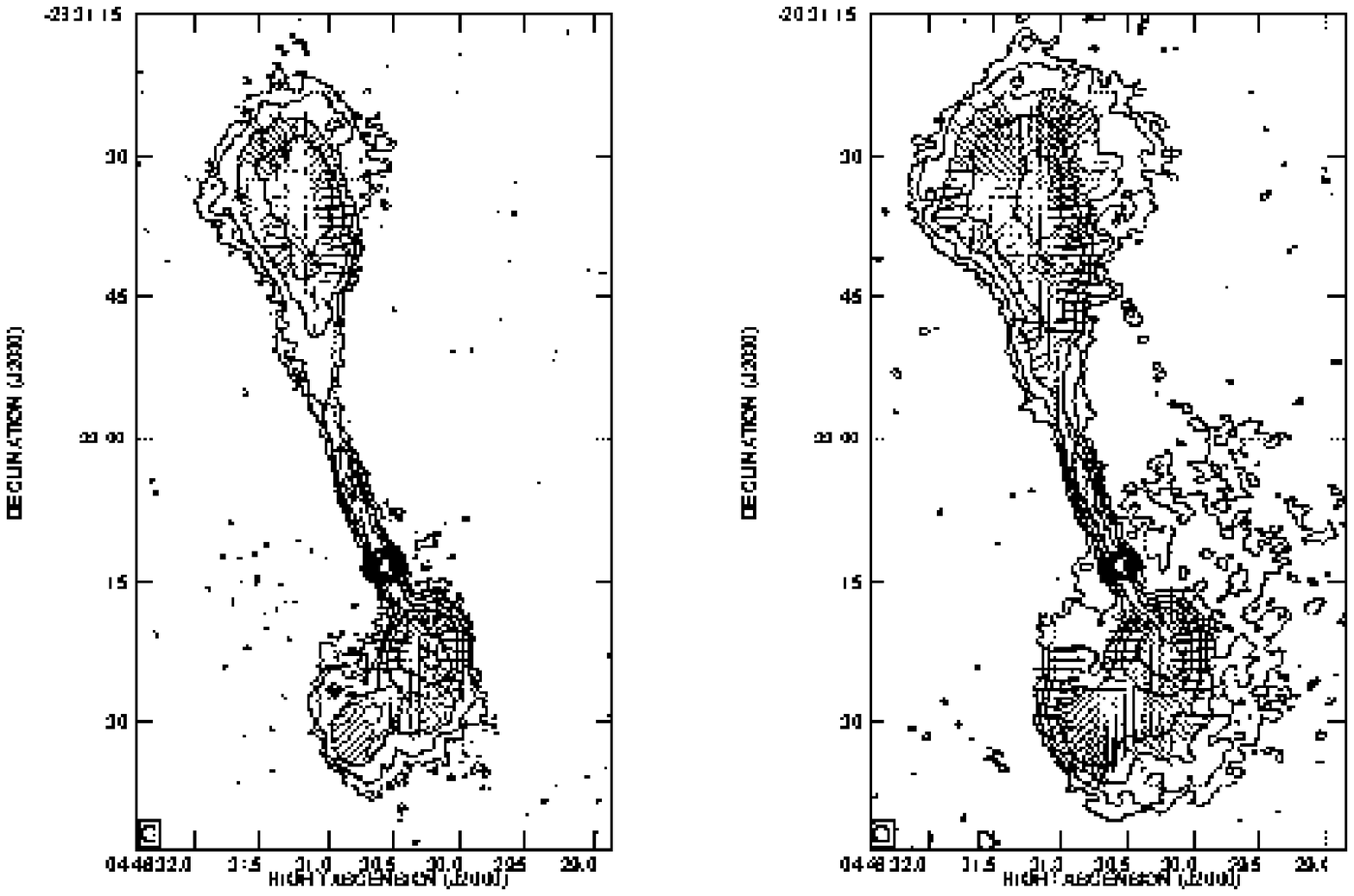}}
\hfill
\caption[]{Left: Image of {\tt J0448-2032} (A514D) 
at 3.6 cm with an angular resolution of $1.6''\times1.6''$.
The rms noise level is 0.018 mJy/beam.
The peak is 7.0 mJy/beam.
Right: Image of {\tt J0448-2032} (A514D) at 6 cm with 
an angular resolution of $1.6''\times1.6''$.
The rms noise level is 0.019 mJy/beam.
The peak is 6.4 mJy/beam.
In both images contour levels are:
$-$0.06, 0.06, 0.12, 0.24, 0.48, 0.96, 1.92, and 3.84 mJy/beam.
Superimposed lines represent the orientation of the
electric vector (E-field) and are proportional in length to the 
fractional polarization ($1''\simeq10\%$).}
\label{fig_A514D}
\end{figure*}

This radio galaxy, belonging to A514 (Owen et al. 1993), is
located in the South-East corner of the cluster, and
appears as an FRI in the high resolution image, but is an extended
tailed radio source (see Fig. \ref{fig_RadX}).
In Fig. \ref{fig_A514D} we show the
images of $J0448-2032$ at 3.6 cm and at 6 cm.  In both 
images the synthesized restoring beam is a circular Gaussian with a
FWHM of 1.6$''$.  The radio images reveal a bright unresolved core
and two oppositely directed jets.  The jet to the North is well collimated up
to about 20 \arcsec where it flares, widening and bending slightly before
diffusing into the Northern lobe.  The jet to the South
is much shorter (a few arcsec), and its emission
blends rapidly into the Southern lobe.  
The structure of the southern lobe suggests a smooth bending 
at a small angle along the line of sight. 
The surface brightness of this lobe, and possibly the Northern 
lobe as well, fades gradually.
The maximum
projected angular size is about 85$''$, and in the radio images
presented here we do not detect the low brightness southern tail
clearly visible in the NVSS image (see Fig. \ref{fig_RadX}) 
and in the image at 1.46 GHz of Owen et al. (1993).
At this frequency, the total flux density calculated by Owen et
al. (1993) is 502 mJy. In our images no polarized emission is 
detected in the core region.
At both wavelengths the jet on the North side has an
average polarization percentage of $\sim7\%$ and the 
percentage of polarization increases along the jet with 
distance from the core.
Although the southern jet is very short
and extends only for about two resolution elements, the jet has
an average polarization percentage of $\sim$5$\%$. 
This last measure is, however, uncertain given that 
the jet is rather short and is embedded in the  
southern lobe emission.
Both lobes are strongly polarized at both 6 and 3.6 cm.
At both wavelengths the lobe on the North side has a 
polarization percentage of about
12$\%$, and the lobe on the South has a polarization percentage of
about 16$\%$.

We found an inverted spectrum  ($\alpha \simeq$ $-$0.1)
in the core and a mean spectral index $\alpha \simeq$ 1.1
in both brighter parts of the lobes.

\subsection{J0448-2044 (A514E)}

\begin{figure*}
\resizebox{12 cm}{!}{\includegraphics {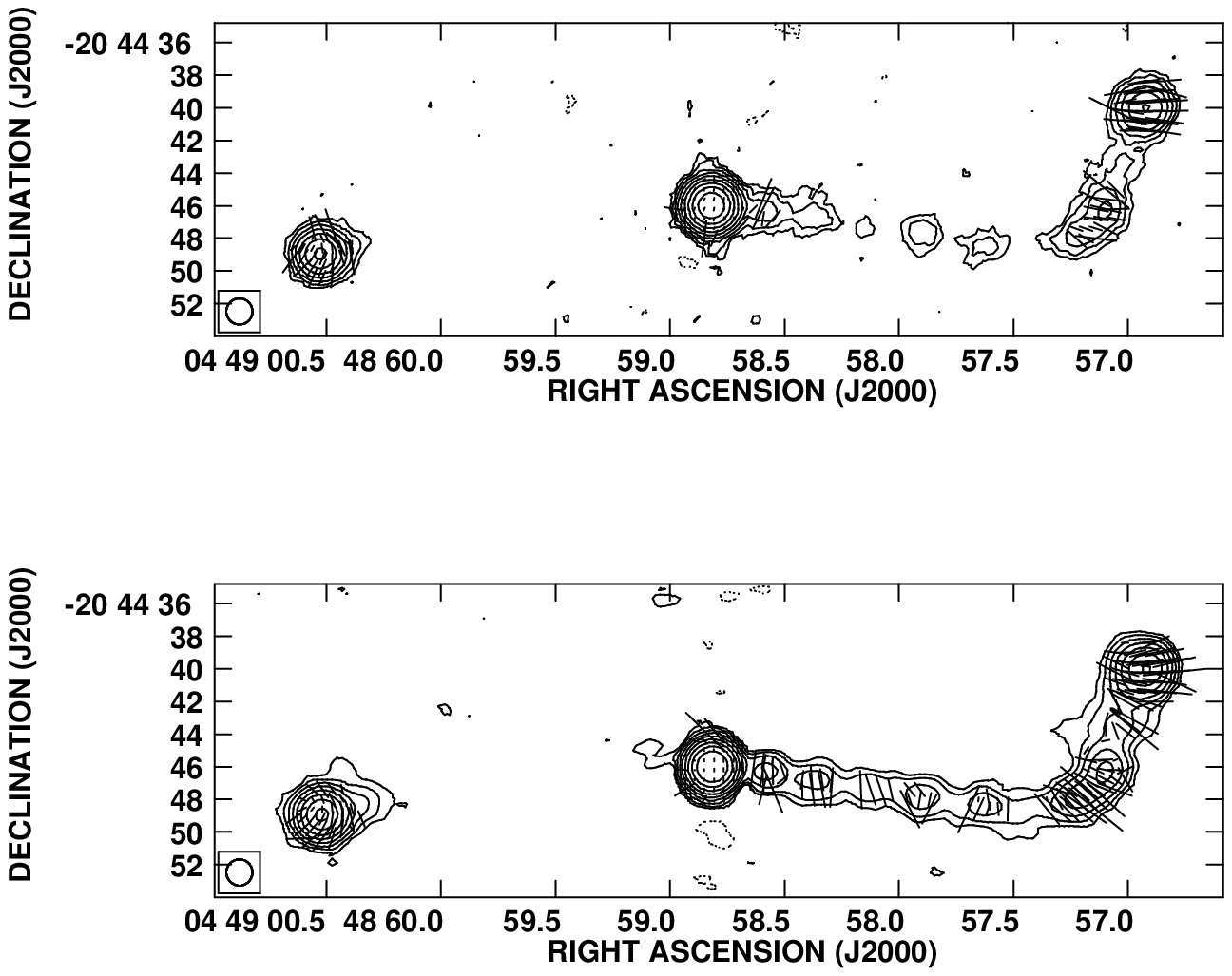}}
\hfill
\caption[]{Top: Image of {\tt J0448-2044} (A514E) at 3.6 
cm with an angular resolution of $1.6''\times1.6''$.
The rms noise level is 0.03 mJy/beam.
The peak is 48.9 mJy/beam.
Bottom: Image of {\tt J0448-2044} (A514E) at 6 cm with 
an angular resolution of $1.6''\times1.6''$.
The rms noise level is 0.026 mJy/beam.
The peak is 65.3 mJy/beam.
In both images contour levels are:
$-$0.1, 0.1, 0.2, 0.4, 0.8, 1.6, 3.2, 6.4, 12.8, and 25.6 mJy/beam.
Superimposed lines represent the orientation of the
electric vector (E-field) and are proportional in length to the 
fractional polarization ($1''\simeq8\%$).}
\label{fig_A514E}
\end{figure*}

This source has been identified with a Quasar at 
a redshift 1.894 (Osmer et al. 1994).
In Fig. \ref{fig_A514E} we show the images at 3.6 cm and at 6 cm.
The maximum angular size of the source is 
$\simeq$55$''$ corresponding to a linear size of about 450 kpc.
A collimated knotty jet starts from the core toward 
the West. At a distance of about 25$''$ from the core the jet bends
to the North.
The hot-spots on the West and East sides are slightly misaligned with
respect to the
core at angles of 21\degrees\ and 18\degrees\ respectively
if we assume that the source axis is defined by the jet.
At the location of the knot in the jet the polarization at 6 cm increases.
The jet on the East side is not visible.

Polarized emission at the level of $\sim$2$\%$ is detected in the core. 
The polarization along the jet is detected at 6 cm and not at 3.6 cm,
probably because of the sensitivity limit.
At 3.6 cm we quote an upper limit of about 20\%
to the fractional polarization.
At the jet bend the percentage of polarization is about 11$\%$ at
both frequencies.
At both wavelengths the hot-spot on the West side  
has a polarization percentage of
about $18\%$,
while the hot-spot on the East    
has a polarization percentage of
about 7$\%$.

We note that the source structure is characterized by the core 
($\alpha \simeq$ 0.5)
and two compact hot-spots ($\alpha \simeq$ 1.2 and 1.3
in the Western and Eastern hot-spots respectively), while the radio lobes
are below our detection limit.

\section{X-ray properties of Abell 514}

From observations with the Einstein Imaging Proportional Counter,
Jones \& Forman (1999), show that the cluster A514 has at least 
three mass condensations with different X-ray luminosities.
A514 was also studied by Ebeling et al. (1996)
in a X-ray flux-limited sample
of 242 Abell cluster of galaxies compiled from the ROSAT All Sky Survey (RASS)
data in the soft X-ray energy band (0.1-2.4 KeV).
They published a X-ray luminosity in the energy band 0.1-2.4 KeV 
of $1.44\times10^{44}$ erg/s and adopted the $kT-L_X$ relation to derive an estimated 
gas temperature of 3.6 keV.

\begin{figure}
\resizebox{9 cm}{!}{\includegraphics {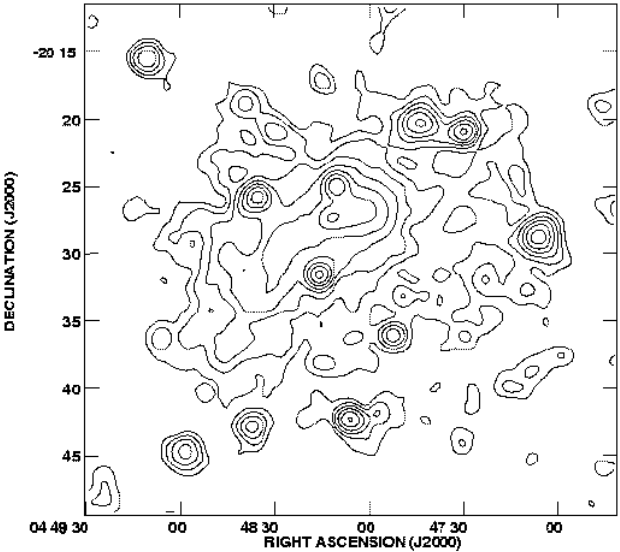}}
\hfill
\caption[]{X-ray PSPC image of the cluster A514.
The image has been smoothed with a Gaussian with $\sigma$=30$''$.
Contour levels are: 0.96, 1.36, 1.92, 2.72,  3.84,  5.43,  7.72, and
10.86 Counts/pixel (1pixel=15\arcsec$\times$15\arcsec).}
\label{A514_X}
\end{figure}

The cluster A514 has been the target of X-ray observations with 
the ROSAT PSPC, for a total exposure time of 18000 seconds.
In Fig. \ref{fig_RadX} (grey scale)
and in Fig. \ref{A514_X} 
(contours) we show the X-ray images in the 0.5-2 keV band 
obtained from the ROSAT public archive
by binning the photon event table in pixels of 15$''$
and by smoothing the image with a Gaussian of $\sigma=30''$.
As already pointed out by Bliton et al. (1998),
the X-ray brightness distribution of A514 is
clumpy and elongated to the South-East direction.

Given its irregular and asymmetric X-ray emission, the cluster
brightness distribution cannot be described by a hydrostatic
isothermal model.
However, in order to estimate the gas parameters 
necessary in the calculation of the magnetic field we attempted
a fit to a portion of the cluster which looks 
approximately regular.

Therefore, part of the X-ray brightness image was fitted with a hydrostatic
isothermal model (Cavaliere \& Fusco-Femiano, 1981):
\begin{equation}
S(r)=S_0(1+r^2/r_c^2) ^{-3\beta+0.5}+S_b
\end{equation}
where S$_0$ is the central surface brightness, r$_c$ is the core radius,
and $\beta$ is the ratio between the galaxy and the gas temperatures.

\begin{figure}
\includegraphics[width=9cm]{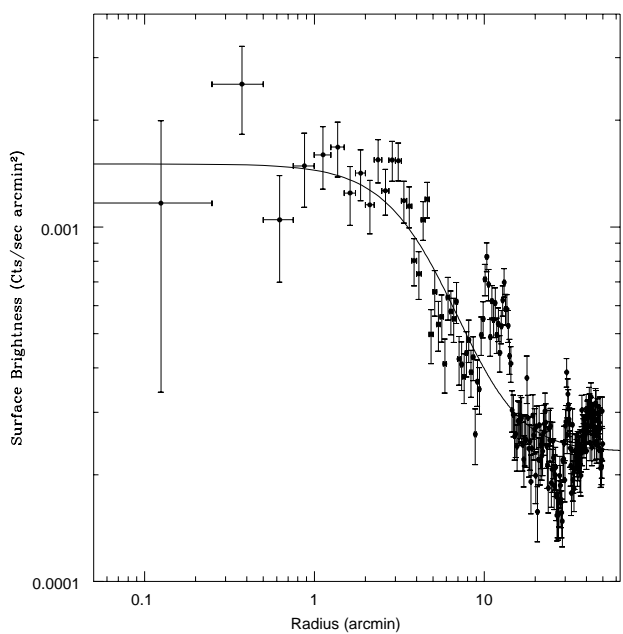}
\hfill
\caption[]{Fit obtained from the surface brightness profile after
subtracting all the discrete sources and considering
only the X-ray emission from North to South on the West side of the
centroid.}
\label{A514_profX}
\end{figure}

It is important to note that due to the cluster irregularity,
the obtained $\beta$-model parameters are strongly dependent on the 
choice of the centroid and the portion of the X-ray surface 
brightness used in the fit.
Different attempts were carried out to test
reasonable values of the X-ray parameters.
In Fig. \ref{A514_profX}  we show the radial
profile of the X-ray surface brightness 
obtained by integrating the PSPC counts over concentric 
rings of $15''$ in radius, after subtracting discrete 
X-ray sources and considering
only the X-ray emission from North to South on the West side of the
centroid. In this way we have excluded from the calculation all the
irregular X-ray extension in the South-East side.
As discussed in Sect. 2, the centroid of the X-ray emission
was taken at the position of the X-ray peak 
RA(2000)=04$^h$48$^m$13$^s$, DEC(2000)=$-$20\degrees 27\arcmin 18\arcsec.
This is approximately coincident with one of the X-ray peaks given by
Jones \& Forman (1999). 

The best fit, reported in Fig. \ref{A514_profX} as a solid line, 
only provides a rough approximation to the data
(reduced $\chi ^2$ $\simeq$ 3.9).
From this fit we obtain a core radius r$_c$=$570^{+200}_{-130}$ kpc and 
$\beta$ =$0.6^{+0.12}_{-0.07}$. The large uncertainty (at 
one sigma level of significance) obtained for the
parameters reflects the irregular cluster structure. 

Assuming a cluster temperature of 3.6 keV
and varying the $\beta$-model parameters within their
interval of confidence we obtain a central density
in the range  0.44-0.50 $\times$10$^{-3}$ cm$^{-3}$.

\section{Rotation Measure in A514}    

\begin{figure}[h]
\resizebox{9 cm}{!}{\includegraphics {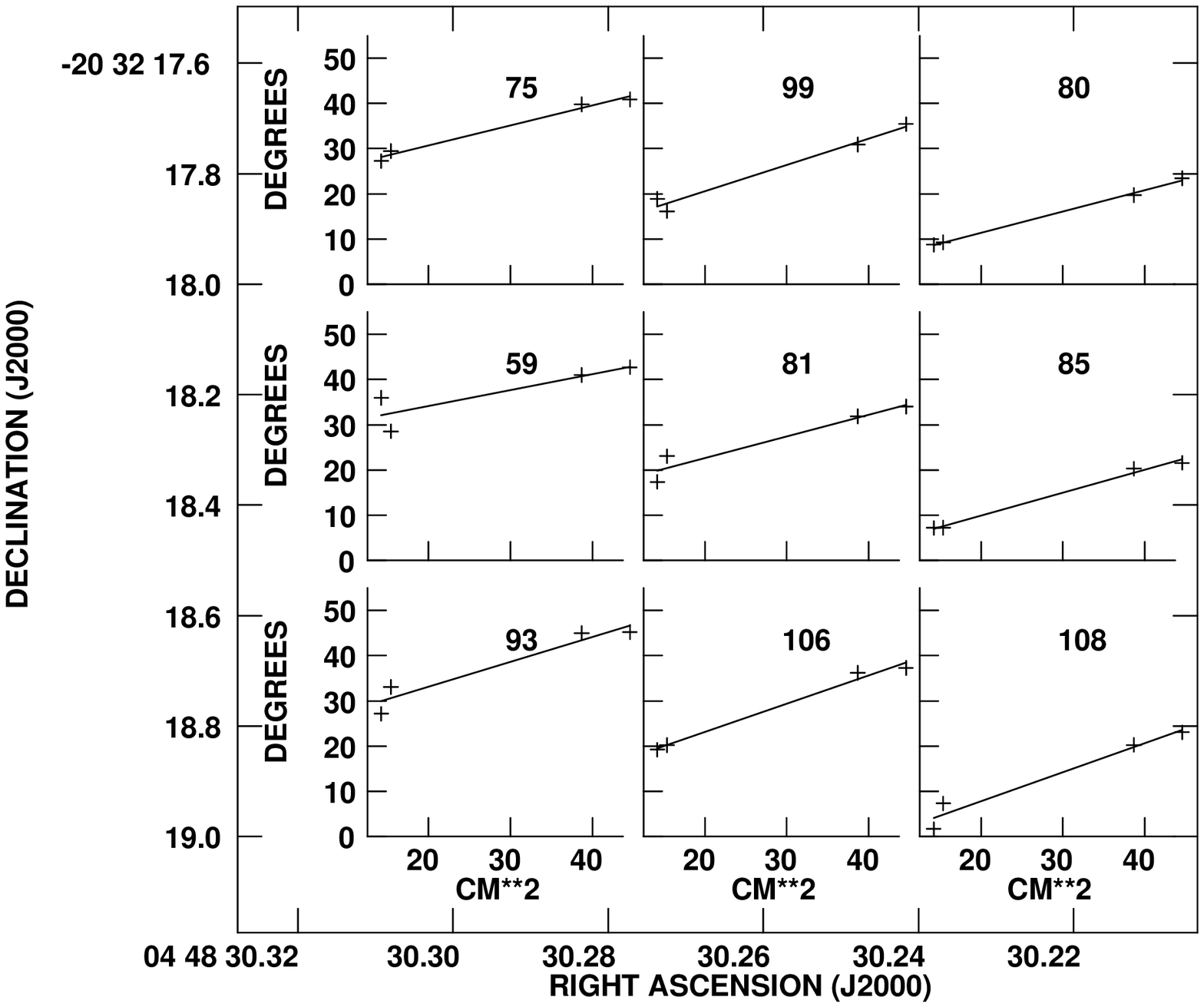}}
\hfill
\caption{The electric vector position angle as a function of $\lambda^2$
and the derived RM fits for some locations in the radio source 
$J0448-2032$ (A514D).
This is an example of how the RM values have been derived throughout.}
\label{RM_fit}
\end{figure}

Polarized radiation from cluster and background radio galaxies 
may be rotated by Faraday effect if magnetic fields are present
in the Intra Cluster Medium (ICM) together with substantial
ionized material.
In this case, the observed polarization angle ($\Psi_{\lambda}$)
is related to the intrinsic polarization angle ($\Psi_{int}$)
through:

\begin{equation}
 \Psi_{\lambda} = \Psi_{int} + RM \lambda^2 
\label{rm}
\end{equation}

where the Rotation Measure (RM) is related to the electron 
density, $n_e$ and magnetic field along the line of sight $H_{\parallel}$
through the cluster according to:

\begin{equation}
RM=812\int_{0}^{L}n_eH_{\parallel}dl~~ {\rm rad/m}^2
\label{equaz}
\end{equation}
where $L$ in kpc, is the path-length through the ICM,
$H_{\parallel}$ is in $\mu$G and 
 $n_e$ in cm$^{-3}$.

The observed RM in a radio galaxy is the sum of 
the contributions to the RM from all magneto-ionic components 
along the line of sight. The RM distribution of radio sources
can therefore be used to derive
information on the magnetic field along line of sights
crossing different regions of the cluster.   

We derived the images of the rotation measure
using the position angles obtained 
at the 4 frequencies 4535, 4885, 8085 and  
8465 MHz with a resolution of 1.6$''$.
Following the definition in Eq. \ref{rm},
the Faraday RMs were obtained by performing a 
least-squares fit of the polarization angle images at each pixel
as a function of $\lambda^2$.
The pixels in which the uncertainty in the polarization
angle exceeds $10^{\circ}$ were blanked.
For each target, except A514B1 where the polarized emission is too weak, 
an image of the RM was obtained. 

\begin{figure*}
\resizebox{18 cm}{!}{\includegraphics {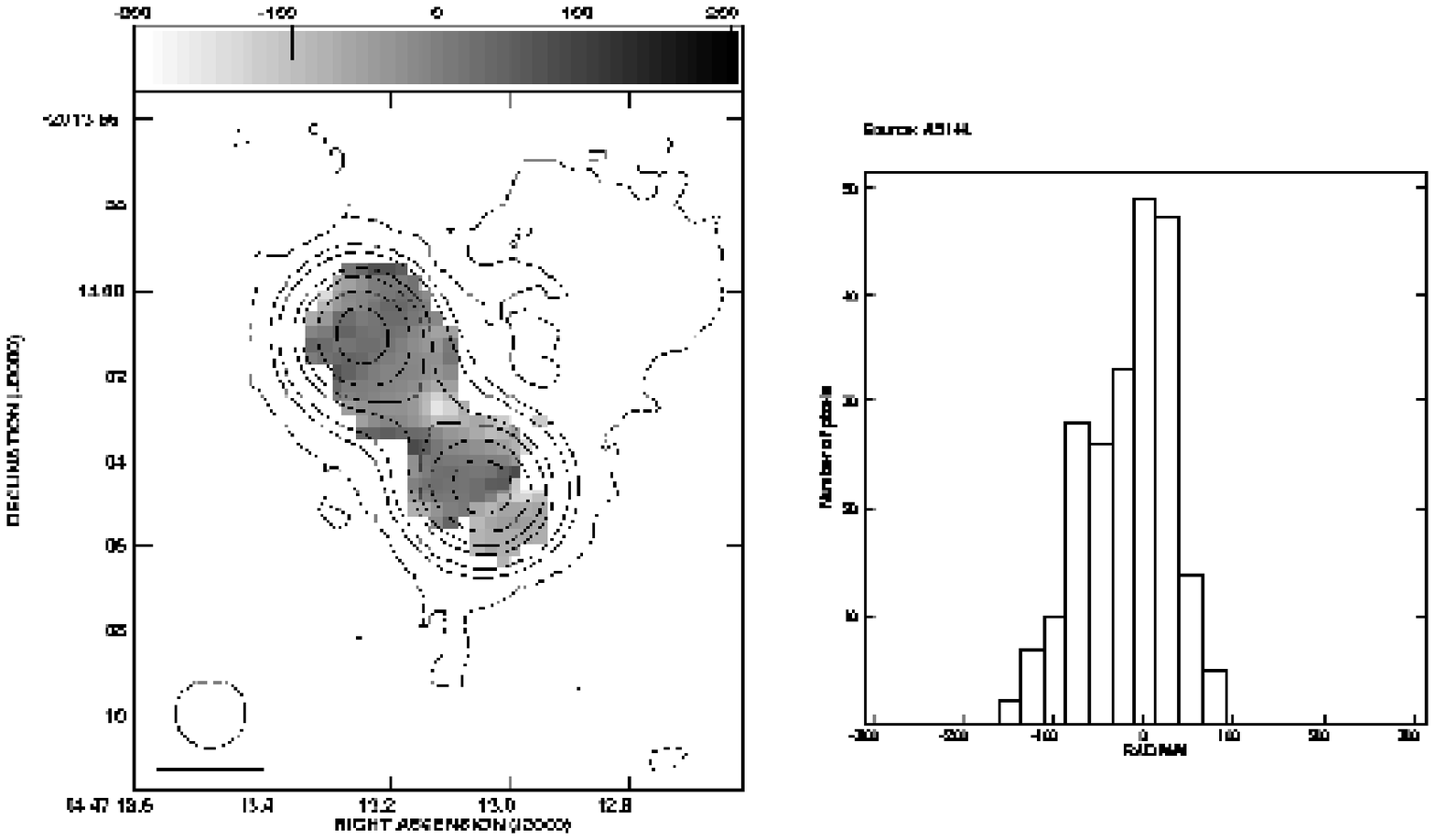}}
\hfill
\caption[]{Left: The image of the rotation measure in 
A514A ({\tt J0447-2014}), 
computed using the polarized angle images at
the frequencies 4535, 4885, 8085 and 8465 MHz with a resolution of 1.6''.
The contours refer to the total intensity image at 3.6 cm.
Right: The histogram of the rotation measure 
for all significant pixels.
The values of RM range between $-$150 rad/m$^2$ and 100 rad/m$^2$. 
The $<$RM$>$ is $-$19 rad/m$^2$
and the $\sigma_{RM}$ is 48 rad/m$^2$.
}
\label{A514A_RM}
\end{figure*}

\begin{figure*}
\includegraphics[width=18cm]{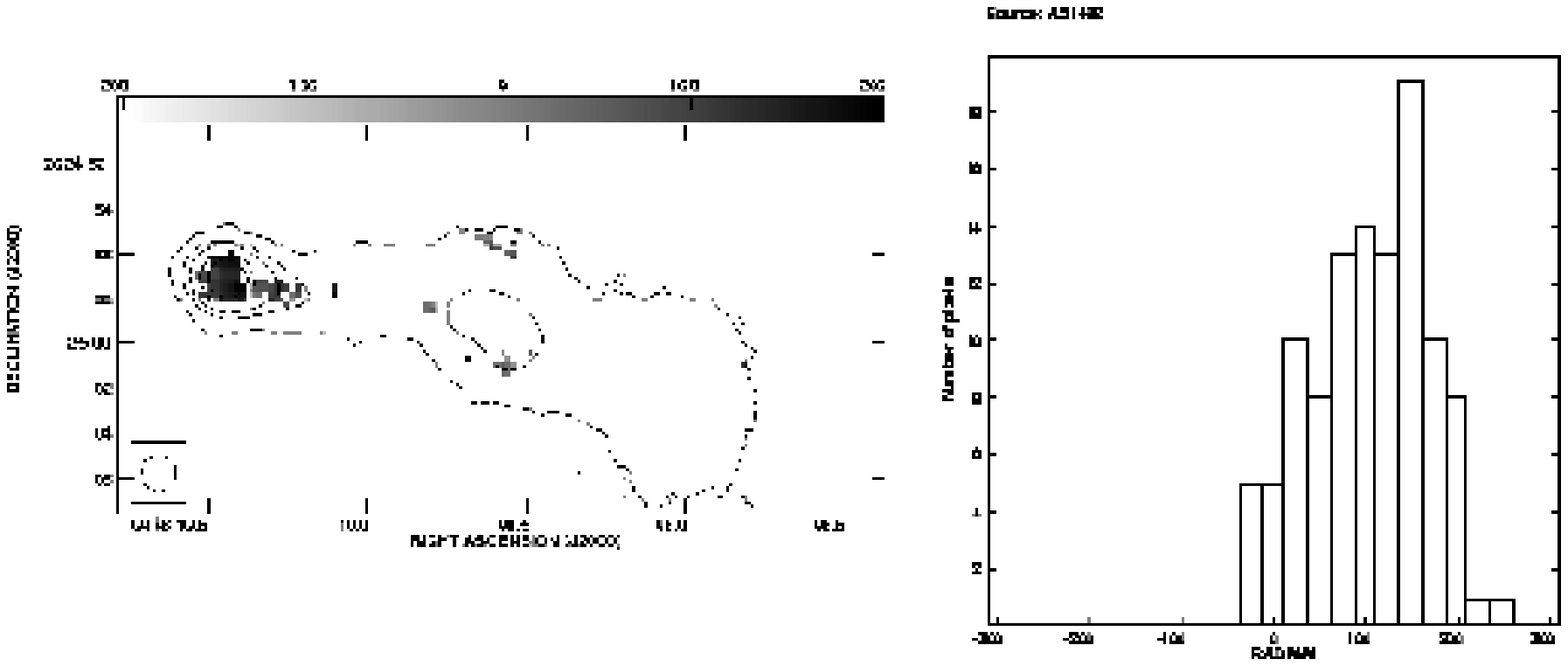}
\hfill
\caption[]{Left: The image of the rotation measure in A514B2 
({\tt J0448-2025}), computed using the polarized angle images at
the frequencies 4535, 4885, 8085 and 8465 MHz with a resolution of 1.6''.
The contours refer to the total intensity image at 3.6 cm.
Right: The histogram of the rotation measure 
for all significant pixels.
The values of RM range between $-$50 rad/m$^2$ and 250 rad/m$^2$. 
The $<$RM$>$ is 104~rad/m$^2$
and the $\sigma_{RM}$ is 63~rad/m$^2$.}
\label{A514B2_RM}
\end{figure*}

\begin{figure*}
\resizebox{18 cm}{!}{\includegraphics {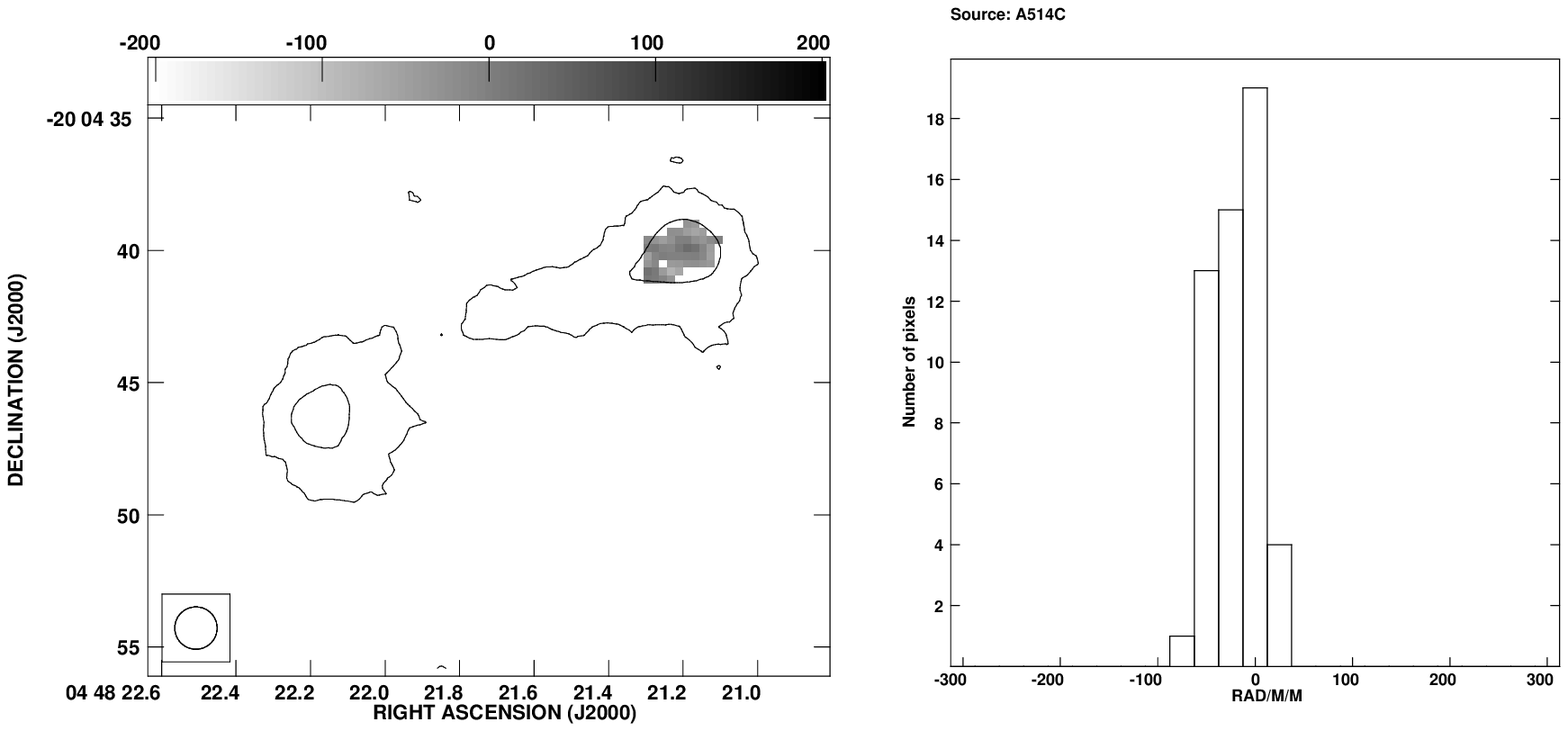}}
\hfill
\caption[]{Left: The image of the rotation measure 
in A514C ({\tt J0448-2005}), computed using the polarized angle 
images at the frequencies 4535, 4885, 8085 and 8465 MHz 
with a resolution of 1.6''.
The contours refer to the total intensity image at 3.6 cm.
Right: The histogram of the rotation measure 
for all significant pixels.
The values of RM range between $-$80~rad/m$^2$ and 50~rad/m$^2$.
The $<$RM$>$ is $-$19~rad/m$^2$
and the $\sigma_{RM}$ is 24~rad/m$^2$.
}
\label{A514C_RM}
\end{figure*}

\begin{figure*}
\resizebox{18 cm}{!}{\includegraphics {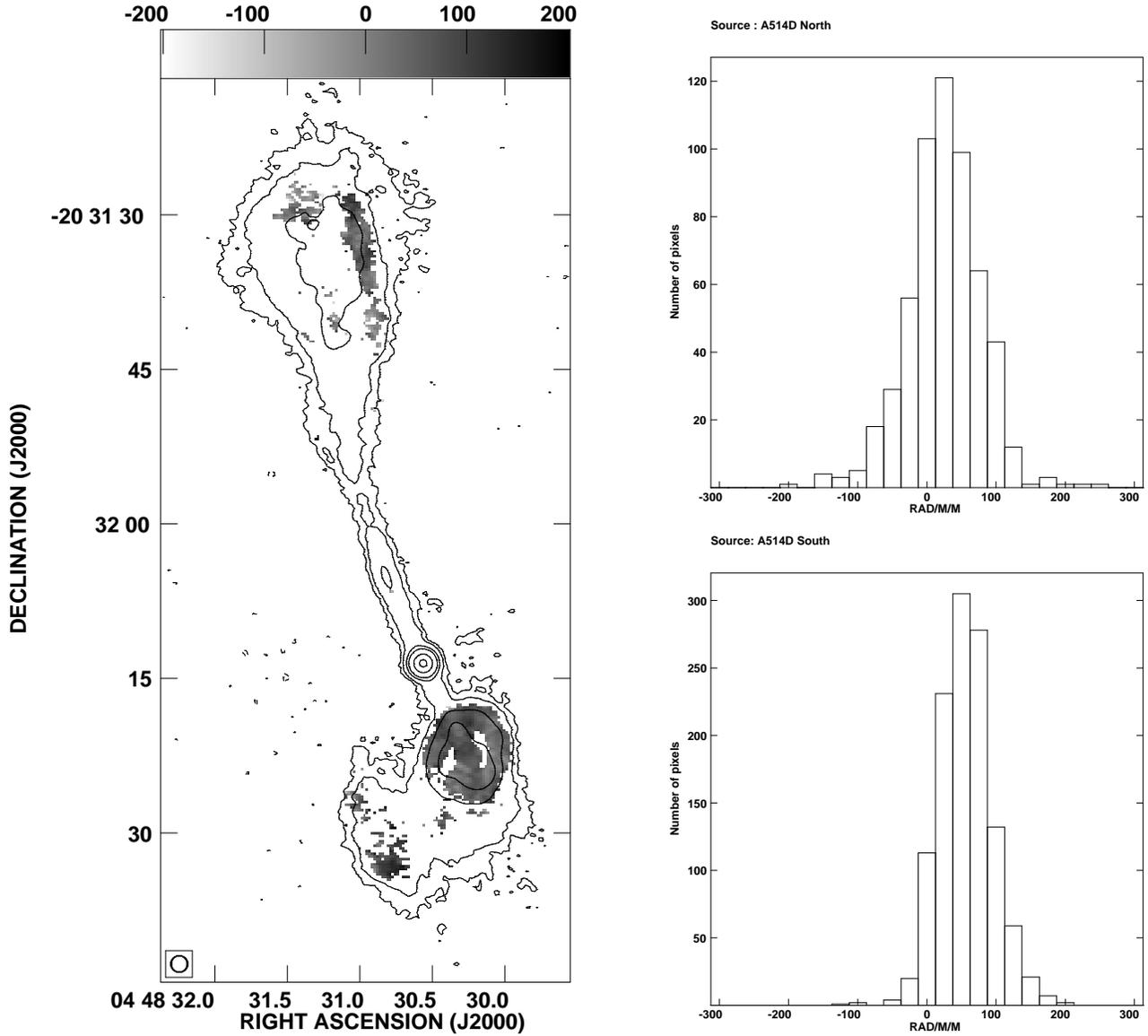}}
\hfill
\caption[]{Left: The image of the rotation measure 
in A514D ({\tt J0448-2032}), computed using the polarized angle images at
the frequencies 4535, 4885, 8085 and 8465 MHz with a resolution of 1.6''.
The contours refer to the total intensity image at 3.6 cm.
Right: The histograms of the rotation measure 
for all significant pixels for North and the South lobe
respectively. The $<$RM$>$ and $\sigma_{RM}$ for the North and the South lobes
are reported in Table \ref{tabella}, while considering all
the significant pixels, the values of RM range between 
$-$200~rad/m$^2$ and +240~rad/m$^2$. 
}
\label{A514D_RM}
\end{figure*}

\begin{figure*}
\resizebox{18 cm}{!}{\includegraphics {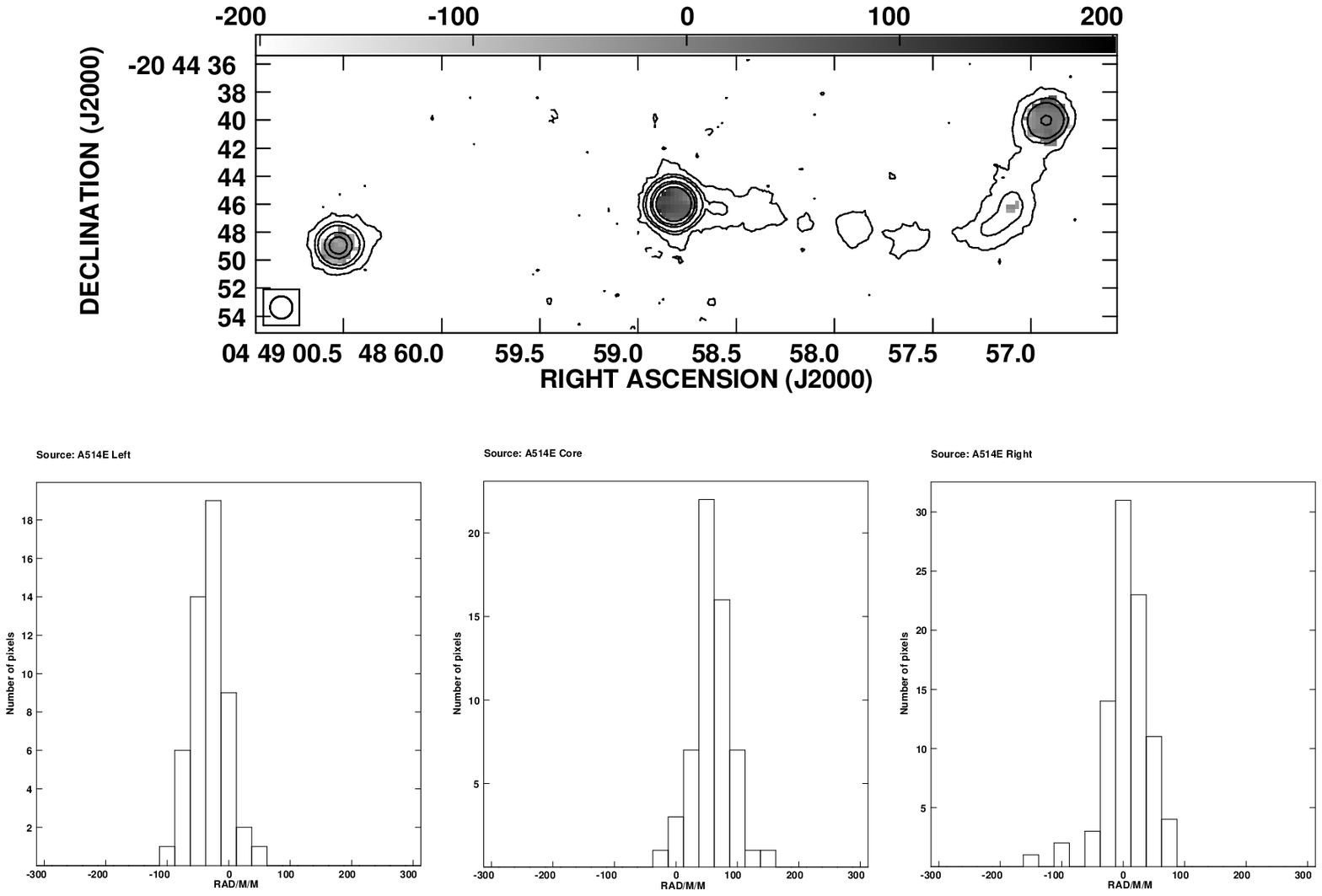}}
\hfill
\caption[]{Top: The image of the rotation measure in 
A514E ({\tt J0448-2044}), computed using the polarized angle images at
the frequencies 4535, 4885, 8085 and 8465 MHz with a resolution of 1.6''.
The contours refer to the total intensity image at 3.6 cm.
Bottom: The histograms of the rotation measure 
for all significant pixels for the East lobe, core and the West lobe
respectively. The $<$RM$>$ and $\sigma_{RM}$ for the East lobe, 
core and the West lobe are reported in Table \ref{tabella}, 
while considering all the significant pixels, 
the values of RM range between $-$150~rad/m$^2$ and $+$150~rad/m$^2$. 
}
\label{A514E_RM}
\end{figure*}

In Fig. \ref{RM_fit} we give examples of RM fits in the source A514D.
The images and the histograms of the RM distribution are presented in 
Figs. \ref{A514A_RM}, \ref{A514B2_RM}, \ref{A514C_RM}, 
\ref{A514D_RM} and \ref{A514E_RM}. 
In each caption we give a brief description of the RM distribution.

\begin{figure}
\resizebox{9 cm}{!}{\includegraphics {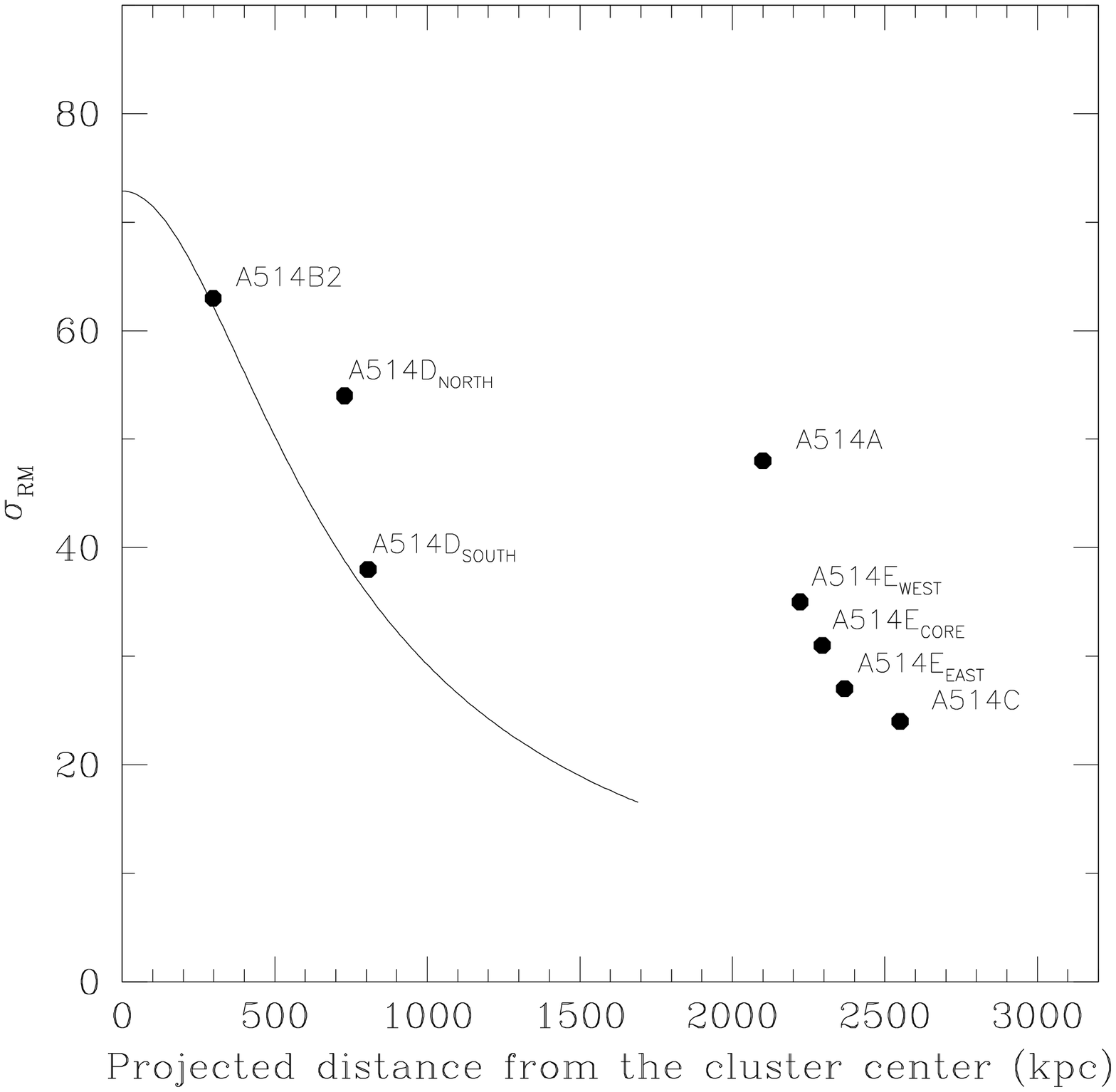}}
\hfill
\caption[]{Values of $\sigma_{RM}$ versus 
the distance from the X-ray cluster center.
The solid line shows the radial profile of $\sigma_{RM}$
as expected from Eq. \ref{rm2}, up to a distance covering the
detected cluster X-ray emission.}
\label{A514_rm}
\end{figure}

\begin{table*}
\caption{Rotation Measure}
\begin{flushleft}
\begin{tabular}{cccccc}
\hline 
\noalign{\smallskip}
Name &   Dist. cluster center & $\mid$RM$_{max}\mid$& $<RM>$     & $\sigma_{RM}$  & Cluster member \\
     &        $(Mpc)$      & $({\rm rad/m}^2)$       & $({\rm rad/m}^2)$  & $({\rm rad/m}^2)$    &  \\
\noalign{\smallskip}
\hline
\noalign{\smallskip}
 A514B2                    &  0.30  & 154 & 104    & 63 & y  \\
A514D (North)              &  0.73  & 120 &  25    & 54 & y  \\
A514D (South)              &  0.81  & 126 &  56    & 38 & y  \\
 A514A                     &  2.10  & 105 & $-$19  & 48 & n  \\
 A514E  (right lobe)       &  2.22  &  34 &  6     & 35 & n  \\
 A514E  (core)             &  2.30  &  68 & 58     & 31 & n  \\       
 A514E  (left lobe)        &  2.36  &  26 & $-$31  & 27 & n  \\
 A514C                     &  2.55  &  15 & $-$19  & 24 & n  \\
\hline
\noalign{\smallskip}
\multicolumn{5}{l}{\scriptsize Col. 1: Source name; 
Col. 2: Distance from the cluster center (arcmin); Col. 3: Maximum absolute value of RM;}\\ 
\multicolumn{5}{l}{\scriptsize Col. 4: Average value of RM; Col. 5: RM dispersion; Col. 6: Cluster member.}\\
\label{tabella} 
\end{tabular}
\end{flushleft}
\end{table*}

In galactic coordinates A514 is located at $l$=219\degrees and 
$b$=$-36$\degrees.
The Galactic contribution to the RM in the region of A514 is expected to be
about $-$16 rad/m$^2$ based on the average of the RM galactic contribution
published by Simard-Normandin et al. (1981) for sources near the cluster.
The aforementioned value is consistent with the RM we derive
for the sources located in the cluster periphery. 
Hereafter we do not apply the correction of the galactic RM since
we are most interested in the differential RM at various cluster
locations.

In the Table \ref{tabella} 
we report the maximum absolute value, the mean, 
and the $\sigma$ of the RM values for the sources
ordered according to an increasing projected distance from the X-ray 
cluster center.
We expect a contribution to the $\sigma_{RM}$ from the noise in the 
measurements at a level of $\sim$10 rad/m$^2$.

Table \ref{tabella} shows that the innermost sources 
have the largest $\sigma_{RM}$
and highest absolute  RMs, and that these values fall
with increasing projected distance from the cluster center.
This result is consistent with the interpretation that the external Faraday 
screen can be the same for all 5 sources, {\it i.e.}, 
the radial profile of $\sigma_{RM}$ in Fig. \ref{A514_rm}
is due to the intracluster medium in A514,
whose differential contribution depends on how much
magneto-ionized medium is crossed by the polarized emission.
Thus, the data show good evidence for the existence of magnetic fields
associated with the intracluster medium.


The RM images 
show fluctuations on scales of 
about 5$''$ (9 kpc at the cluster redshift), 
therefore we can consider this value as the coherence 
length of the magnetic field.
This scale can be primarily evaluated from the RM structure across
the stronger and more extended source $J0448-2032$ (A514D)
and this value is consistent with
coherence length of the magnetic field found 
in previous works in the literature see e.g. Coma (Feretti et al. 1995), 
A119 (Feretti et al. 1999) and 3C129 (Taylor et al. 2001).

\section{Magnetic field in A514}

The determination of the strength of the magnetic field in a cluster 
depends on several assumptions, including the model for the X-ray
emitting gas distribution and the magnetic field structure.

The effect of a Faraday screen with
different gas density distributions and with a tangled magnetic 
field has been analyzed by several authors
(Lawler \& Dennison 1982, Tribble 1991, Felten 1996).
Assuming a randomly
oriented magnetic field
in cells of uniform size and strength, and a gas 
density distribution given by the hydrostatic isothermal $\beta$ model
(Cavaliere \& Fusco-Femiano 1981) {\it i.e.}:
\begin{equation}
n_e(r) =n_0 (1 + r^2/r^2_{\rm c})^{-3 \beta/2}
\end{equation}
where $n_0$ is the central (electron) density, and $r_c$
is the core radius of the gas distribution,
the RM dispersion at 
different projected distances from the cluster center
was evaluated by Feretti et al. (1995) and Felten (1996) 
by solving the integral of Eq. \ref{equaz}:
\begin{equation}
 \sigma_{RM}= {{441 H n_0  r_c^{1/2} l^{1/2} }\over
{(1+r^2/r_c^2)^{(6\beta -1)/4}}} \sqrt {{\Gamma(3\beta-0.5)}\over{\Gamma
(3\beta)}}
\label{rm2}
\end{equation}
where $\Gamma$ is the Gamma function, 
$r_c$ is the core radius in kpc, and $l$ is the size of
each cell in kpc,  the central gas density $n_0$ is in cm$^{-3}$ 
and $H$ is in $\mu$G; and it is assumed that the source
is as distant from the observer as the cluster center.

For the cluster galaxies $J0448-2025$ (A514B2) and $J0448-2032$ (A514D),
using the gas parameters calculated from the $\beta$-model fit,
the values of $\sigma_{RM}$ 
fit fairly well in the model described above and 
are consistent with a magnetic field strength in
the range 3-7 $\mu G$ for a cell size of about 9 kpc.
The uncertainty in the magnetic field strength
is dominated by errors in the gas parameters 
calculated from the $\beta$-model fit.
Assuming the above magnetic field, the cell size, and 
the $\beta$-model fit calculated for A514, we derived
the radial profile
of $\sigma_{RM}$ expected from Eq. \ref{rm2}. This is
plotted as a solid line in Fig. \ref{A514_rm}, up to
a distance corresponding to the extent of the cluster X-ray emission
above the background.

The calculated $\sigma_{RM}$ of the background 
sources (A514A, A514C, A514E) is generally higher than
that expected from the absence of cluster contribution at these
large projected distances from the cluster center.
This may be due effects internal to the radio sources,
to local effects of the host galaxies or to some other
effects along the line of sight.

The model we used to derive the magnetic field strength
is oversimplified and has the following
limitations:
(1) the radio galaxies belonging to the
cluster can be at different locations
along the line of sight; (2) the cluster shows an irregular
X-ray emission, therefore the described King density model is
only a rough approximation; and (3) 
the magnetic
field structure is likely to be more complicated than assumed.

Even allowing for the uncertainties related to the previous computation,
the observational evidence favors
the existence of a strong magnetic field  in the
intergalactic medium of A514, over a wide scale
of about 1.4 Mpc in diameter.   

At the cluster center the energy density in the magnetic
field results about $4-24\%$ of the thermal energy density.

\section{Conclusions}
We have obtained indirect evidence for the presence of a 
magnetic field in the cluster A514.
Six radio sources 
located at different projected distances from the cluster
center of A514 have been studied in total intensity and polarization.  
In conjunction with hot gas density estimates based on the X-ray properties
of the cluster, observations of Faraday rotation measures in the radio
sources can most reasonably be explained by the presence of cluster 
magnetic fields with a strength of $\sim$3-7 $\mu G$ spread throughout
the central 1.4 Mpc of the cluster.
This magnetic field is consistent with the magnetic
field calculated using similar methods in other clusters of galaxies 
without a cooling flow (see e.g. Coma, A119, 3C129).

\section*{Acknowledgments} 
This work was partly supported by the Italian Ministry for University
and Research (MURST).
This research has 
made use of the
NASA/IPAC Extragalactic Data Base (NED) which is operated by the JPL, 
California Institute of Technology, under contract with the National 
Aeronautics and Space Administration.

\end{document}